\documentclass[preprint2]{aastex}
\shorttitle{Southern Proper Motion Program III}

\newcommand{\be}{\begin{equation}}
\newcommand{\ee}{\end{equation}}

\begin{document}
\input{psfig.sty}

\title{The Southern Proper Motion Program III. \\
    A Near-Complete Catalog to V=17.5}


\author{Terrence M. Girard, Dana I. Dinescu, William F. van  Altena}
\affil{Astronomy Department, Yale University, P.O. Box 208101,
       New Haven, CT 06520-8101}
\email{girard@astro.yale.edu}

\author{Imants Platais}
\affil{The Johns Hopkins University, Department of Physics and Astronomy,
3400 N. Charles St., Baltimore, MD 21218}

\author{David G. Monet}
\affil{US Naval Observatory, Flagstaff Station, P.O. Box 1149, 
Flagstaff, AZ 86002}

\author{Carlos E. L\'{o}pez}
\affil{Universidad de San Juan and Yale Southern Obs., 
Av. Benav\'{i}dez 8175 Oeste, Chimbas, 5413 San Juan, Argentina}

\begin{abstract}

We present the third installment of the Yale/San Juan Southern Proper
Motion Catalog, SPM3.  
Absolute proper motions, positions, and photographic B,V photometry
are given for roughly 10.7 million objects, primarily stars, down to a
magnitude of V=17.5.
The Catalog covers an irregular area of 3700 square degrees, between
the declinations of -20$\degr$ and -45$\degr$, excluding the Galactic plane.
The proper-motion precision, for well-measured stars, is estimated to
be 4.0 mas~yr$^{-1}$.
Unlike previous releases of the SPM Catalog, the proper motions are on the
International Celestial Reference System by way of Hipparcos Catalog stars,
and have an estimated systematic uncertainty of 0.4 mas~yr$^{-1}$.
The SPM3 Catalog is available via electronic transfer.
\footnote{http://www.astro.yale.edu/astrom/}

As an example of the potential of the SPM3 proper motions, we examine the
Galactocentric velocities of a group of metal-poor, main-sequence A stars.
The majority of these exhibit thick-disk kinematics, lending support
to their interpretation as thick-disk blue stragglers, as opposed to
being an accreted component.

\end{abstract}


\section{Introduction}


The Yale/San Juan Southern Proper Motion (SPM) Program was originally
intended as a southern-sky complement to the Lick Northern Proper
Motion (NPM) Program (Klemola et al.~1987).
Photographic B,V photometry, positions and absolute proper motions would
be determined for a significant fraction of the southern sky, as
outlined by van Altena et al.~(1990, 1994).
The photographic plates, upon which the SPM is based, extend to
approximately V=18.
Previous versions of the SPM Catalog were derived from PDS microdensitometer
measures of these plates.
Limited by the throughput of the PDS system, inclusion of all measureable
images on the plates was impossible.
Instead, an input list was compiled, consisting of stars of special
interest gleaned from the literature, all measureable galaxies (needed
for the absolute proper-motion reference frame), and a randomly selected
sample of stars intended for kinematic study.

Such an input list was used to construct the first SPM Catalog which covered
an area of 720 square degrees at the South Galactic Pole (SGP) and included
roughly 50,000 stars and 9,000 galaxies, (Platais et al.~1998, 
hereafter Paper II).
The second SPM Catalog extended this area to a 3700-square-degree band 
in declination from approximately -20$\degr$ to -45$\degr$, but excluding the 
Galactic plane fields.
This second catalog, the SPM2
\footnote{described in detail at http://www.astro.yale.edu/astrom/},
also included a re-reduction of the measures
that comprised the SPM1 Catalog.
The SPM2 contained 287,000 stars and 35,000 extragalactic objects.
This represents less than 3\% of the total number of measureable stars on
these plates.

The present catalog, SPM3, is based on the same plates and covers
the same area as the SPM2.
However, the plates have now been rescanned using the Precision Measuring 
Machine (PMM) of the US Naval Observatory, Flagstaff Station, 
(Monet et al.~2003 includes a description).
This allows all identifiable images to be measured, albeit to a slightly
reduced precision relative to that attainable with the Yale PDS.
The result, is a catalog of absolute proper motions, positions, and 
B,V photometry that is approximately complete to V=17.5.
(We estimate the incompleteness due to confusion caused
by overlapping images and unidentified high proper-motion stars to be 5\%.)
The SPM3 contains a total of 10.7 million objects,
over the same 3700 square degrees covered by the SPM2. 

As with earlier versions of the SPM Catalog, an explicit correction has
been made for magnitude equation -- the magnitude-dependent systematic
shift in image position that afflicts all photographic material to
some extent.
Among the various photographically based proper-motion catalogs
currently available, the SPM is unique in this important regard.
Magnitude equation, and its correction, is discussed in detail by
Girard et al.~1998, (hereafter, Paper I).
Whereas earlier versions of the SPM Catalog used external galaxies to
set the zero-point of the absolute proper-motion frame, the SPM3 is
on the proper-motion system of the International Celestial Reference System
(ICRS) via the Hipparcos Catalog (ESA 1997).

In the following section, the plate material and plate measurement are
described.  
In Section 3, the astrometric and photometric reduction procedures and
catalog construction are detailed.  
In Section 4, the properties of the catalog are examined by means of
internal and external comparisons.  
The various incarnations of the SPM Catalog are described in Section 5,
to assist the potential user in selecting the appropriate version.
Section 6 illustrates a practical application of the SPM3 Catalog -- using the
kinematic data to better determine the nature of a group of
metal-poor, main-sequence A stars, first pointed out by Rodgers (1971).
The final section summarizes the current status and future direction
of the SPM program.

\section{Plate Material and Measurement}


Observations were carried out using the 51-cm double astrograph of
the Cesco Observatory in El Leoncito, Argentina.
The double astrograph is essentially two telescopes sharing a common tube.
One of the objective lenses is optimized for the visual passband, the
other for blue.
Originally, the SPM program was to extend from the south celestial pole to
the equator, providing substantial overlap with the NPM program that
extends from the north down to roughly -23$\degr$ declination.
Due to limited resources, the first epoch of the SPM program was halted
after coverage from the southern celestial pole to about -20$\degr$
was completed, roughly 76 percent of the southern hemisphere.
The first-epoch plates were taken in the late 1960's to early 1970's.
The second-epoch survey was begun in the late 1980's, but was only about
35 percent completed when Kodak discontinued production of the 103-a emulsions.
Most of the SPM fields for which second-epoch plates exist have been included
in the SPM3, with the exception of the low Galactic-latitude fields.
The SPM3 Catalog is based on 156 such SPM fields.
SPM sky coverage is illustrated in Fig.~1.
The mean epoch of the first-epoch plates is 1968.0 and that of the
second-epoch is 1991.4.

With a scale of $55\farcs1$ mm$^{-1}$, the 17-inch photographic
plates used cover an area of 6.3$\degr$ $\times$ 6.3$\degr$.
They are taken on 5$\degr$ centers in declination, with varying offsets
in right ascension, but never exceeding 5$\degr$.
In most cases, the blue and visual plates for a specific field were
exposed simultaneously.
Each plate contained two exposures, one of 2 hours duration, and an offset
exposure of 2 minutes duration.
Each of the exposures was centered on the meridian.
Also, an objective grating, with grating constant $\Delta$m=3.85, was used to
produce diffraction-image pairs that bracket the brighter central-order images.
These grating images not only extend the useful dynamic range of the plates
but also provide a means for calibration and correction
of each plate's magnitude equation.

The blue and visual pairs of first and second-epoch plates were scanned
with the Precision Measuring Machine (PMM) of the USNO Flagstaff Station.
We present here only a brief description of the PMM hardware and software.
For a more complete description, see Monet et al.~(2003) and references
therein.

The PMM uses a 2-dimensional CCD detector (1312$\times$1033 useful pixels)
to snap a series of ``footprint'' images of the photographic plate.
The platen, holding the plate, is repositioned precisely before each exposure
in a manner such that the entire 17-inch plate is scanned by an array of
26$\times$34 such footprints.
There is a small amount of overlap between adjacent footprints.
The effective pixel size is $0\farcs75$.

Each footprint, consisting of an array of transmission values, is processed
in real time by the PMM software pipeline.
A ``blob''-detecting routine identifies groups of pixels above the 
locally-determined sky background.
A 5-parameter functional fit is then performed, by least squares, to the
transmission values, $T(x,y)$, minus background, $B_T$.
The circularly symmetric function used in the PMM pipeline is the following,
\be
  T(x,y) - B_T = A [e^{\alpha((x-x_0)^2 + (y-y_0)^2 -r_0^2)} + 1 ]^{-1}
\label{eq1}
\ee
where $A$ is the image's amplitude, $x_0$ and $y_0$ are its position,
and $r_0$ and $\alpha$ represent a saturation radius and a measure of
the wings' extent, respectively.
The values of $x$ and $y$ within the footprint are simply the CCD pixel
coordinates multiplied by the nominal pixel scale.

The position of the CCD at the time the footprint is taken, is measured
accurately by a laser interferometer.
This is combined with the derived image center, ($x_0,y_0$), to yield the
plate coordinates of each image detected and centered.
At the time the SPM plates were scanned, the PMM pipeline made a scalar
correction to the nominal pixel scale in each footprint.
We found this was inadequate and instead determined a general linear
correction, (with relative scale, rotation, and skew terms), per footprint by
minimizing the differences in position for multiply measured stars in
the overlap regions of the footprints.
(Subsequently the PMM pipeline has also been modified to include a
field-dependent ``refocus'' correction based on the overlap images.)

Finally, the PMM software calculates additional image parameters.
These include a magnitude estimate based on the integrated flux
of the pixels used to perform the functional fit.
It is this instrumental magnitude estimate which will be used in our
photometric reduction.

\section{SPM Catalog Construction}


The goal of the SPM3 is to be as complete as possible to a
limiting magnitude determined by the plate material.
The most obvious approach is to construct the catalog from the complete
list of images detected by the PMM software, setting a cutoff at some
specified signal-to-noise threshhold level.
Unfortunately, the presence of the grating images, (in both long and short 
exposures), as well as a large number of spurious detections and multiple
detections, make it difficult to implement such a straightforward approach.

Instead, we have chosen to first construct an input catalog from existing,
external catalogs; namely the Hipparcos Catalog (ESA 1997), 
the Tycho-2 Catalog (Hog et al.~2000),
the UCAC 0.9 (see Zacharias et al.~2000 for a description of the UCAC1),
and the Two Micron All Sky Survey
(2MASS) point-source and extended-source catalogs.
The external catalogs are merged, in the order listed above, to provide
an input catalog that is as complete as possible.

Of course, these input catalogs contain substantial overlap, with almost all
Hipparcos stars being in the Tycho-2 catalog, and the majority of
these, as well as almost all UCAC stars, being contained in the 2MASS catalogs.
For the purposes of preparing our master input catalog, duplicates were
identified by simple positional coincidence 
using a matching tolerance of 1.1 arcseconds.


With the input catalog in hand, the next step is to identify each PMM-detected 
image with an image system (e.g. long-exposure, central order) and a specific 
source from the input catalog.
This is accomplished by forming vector differences between the x,y 
coordinates of the brightest
PMM-measured images and the predicted plate coordinates of Tycho-2 stars,
knowing the plate scale, orientation, and approximate plate-center 
tangent point.
The discrete clusters of points in this vector-difference space make it
possible to identify the various exposure/grating systems, and 
determine their relative separations.
Based on these preliminary image-system identifications
and the predicted plate coordinates of the Tycho-2 stars in the
input catalog, approximate transformations are found relating each
of the image systems to celestial coordinates.
The transformations are applied to each image system and an improved
matching with the Tycho-2 stars is made by position, and from these
matches, a new set of image-system transformations is derived.
The form of the final transformation equations is a general cubic polynomial.

Applying the transformations to all images detected on the plate, a
spatial matching is performed with the full input catalog.
A rather large matching tolerance of 100 $\mu$m (5.5 arcsec) is used.
(A cleaning of improperly matched objects is made later.)
All measured images, through third-order in both long and short exposure,
for each input catalog star
are extracted and saved, along with their input-catalog identification.
Once the measurements have been extracted for all four plates in an SPM
field, (B and V plate pairs at first and second epoch), the photometric
and astrometric processing is performed on a field by field basis.

\subsection{Photometric Reduction}


The PMM software provides an instrumental magnitude index, $p$, for each
image detected and centered.
The calibration of these photometric indices to $B, V$ estimates is a 
four-step process.

First, the indices are made nominally linear with true magnitude by use of the 
grating constant, which is assumed to be field and magnitude independent.
A polynomial, of up to fifth-order, plus magnitude-field terms is adopted 
for the form of the ``linear'' magnitude, $m$.
\begin{eqnarray} \nonumber
   m = a_1 p + a_2 p^2 + a_3 p^3 + a_4 p^4 +a_5 p^5 \\
        + a_6 px +a_7 
\label{eq2}
\end{eqnarray}
The coefficients, $a_{k}$, are determined by least squares using the condition
that the difference between the first-order and central-order magnitude
estimates for all stars, for which both can be measured, should be equal to 
the grating constant.
A value of $m_{(n=1)} - m_{(n=0)} = 3.85$ is adopted for the grating constant.
A separate set of $a_{k}$ are determined in the long and short-exposure 
systems, and then applied to all images in that system.
Note that had a constant term been included in Equation (2), it 
would be indeterminable, as would pure field terms.
Also, any deviation in the actual grating constant from the nominal value
of 3.85 will result in a non-unity scale term in the linearized magnitudes.
All of these terms, which cannot be determined at this stage,
are included in the third step of the photometric reduction process.

The second step in the photometric reduction places the short and long-
exposure photometric indices onto a common system.  
Due to the differences
in the way in which the stellar-image profiles vary across the plates in
these two systems, this transformation allows for the inclusion of up to
cubic field terms, when they are deemed significant.
\begin{eqnarray} \nonumber
   m_{long} - m_{short} = b_1 + b_2 x + b_3 y
             + b_4 x^2 + b_5 xy \\ + b_6 y^2 
             + b_7 x^3 + b_8 x^2 y + b_9 xy^2 + b_{10}y^3
\label{eq3}
\end{eqnarray}
In actuality, the constant term in the above expression is redundant, 
but is included for computational convenience.

After application of the above transformation to the short-exposure images,
the photometric indices of all measured grating orders in both exposures,
$m$, will be on a common system, to within a constant, i.e., not a
function of magnitude.
What remains is to determine the constants, for the various grating and
exposure systems, 
i.e., the magnitude
difference between the central-order and various higher diffraction
orders in both the long and short exposures, and any field dependence these
constants might have.
This is done in the third step of the photometry calibration process,
using calibrating reference stars of known magnitudes.
\be
   m_{ref} = c_{1,k} 
           + c_2 x + c_3 y + c_4 x^2 + c_5 xy + c_6 y^2 + d m_k
\label{eq4}
\ee
where $k=$1 to 8 indicates the grating-order/exposure system, central order
through third order in both the long and short exposure.
The reference star magnitudes were originally taken from three sources --
the Guide Star Photometric Catalog (Lasker et al.~1988);
a $B, V$ CCD sequence taken with the CTIO 0.9-m telescope at the center of
each of our SPM fields, specifically taken for the construction of earlier
versions of the SPM Catalog, (see Paper II);
and Tycho-2 Catalog $B, V$ values, in order to provide the necessary
leverage on the field terms.
After some experimentation, we found that the faint end of the
calibration could be better determined by also including the 
(PDS-measured) SPM2 Catalog
stars as photometric calibrators, providing thousands of reference stars
per plate.
Subsequent to this calibration step, multiple magnitude estimates of the
same star, from different exposure/diffraction systems, are combined
by weighted average using a scheme similar to that detailed in Paper II.
This gives a single $B$ and $V$ estimate per star per epoch.

The first and second-epoch magnitude estimates are averaged
to yield individual $B$ and $V$ estimates per star.
Uncertainties are derived from the scatter in the
magnitude estimates from the various exposure/diffraction systems and
from the two different epochs.

Even with the additional SPM2 photometric reference stars, there remained 
systematic differences as a function of magnitude when compared to the
SPM2 photometry.
The bottom panel in Fig.~2 shows a comparison between SPM2 and
preliminary SPM3 $B$-magnitude estimates, for a sample SPM field.
The differences in $V$ are not shown but are similar to those in $B$.
The large deviations in the magnitude calibrations are
undoubtedly rooted in the raw photometric indices.
There is a significant difference in the forms of the image-fitting functions
used to reduce the PMM measures of the SPM3 and that of the PDS measures
of the SPM1 and SPM2.
As our photometric calibration pipeline was originally designed for the
PDS measures, it is likely the simple application of the same
routines to the PMM measures that accounts for the severe magnitude trends
in the derived photometry and the SPM2 data are to be considered more reliable.

Thus, as a last step in the photometric reduction of the SPM3
Catalog, a final correction was applied to each star's $B$ and $V$ estimate.
This correction was a moving median, as a function of magnitude, of the
differences SPM2 photometry minus SPM3 preliminary photometry, on a 
field by field basis.
For each star in the SPM3 input list, the median difference of the 
50 stars nearest in magnitude
to the target star {\it and} in common with the SPM2 was used as a correction
to the SPM3 magnitude.
The bottom panel in Fig.~3 shows the differences between SPM2 and SPM3
photometry {\it after} the moving median corrections have been applied,
for the same sample field shown in Fig.~2.
The scatter on the bright end is disappointingly large, but the systematic
trends have been effectively removed.
(The diagonal banding in the bottom panel of Fig.~3 is an artifact 
caused by the use of the SPM2 data both to adjust the photometry 
and to assess it, as a function of the corrected photometry itself.)

\subsection{Astrometric Reduction}


The astrometric reduction procedure of the SPM3 differs
from that of previous versions of the SPM Catalog, but effectively
achieves similar results by making use of the SPM2 data to calibrate
the final astrometry.
In the past, relative proper motions were derived by using a large number
of faint, anonymous stars to transform measures from a first-epoch plate
into the system of the same passband second-epoch plate --
a differential method.
The correction to absolute proper motions was then a matter of using
the galaxies' relative proper motions to determine the constant offsets,
field by field.

For logistical reasons, preliminary SPM3 proper motions are derived
directly from differences in equatorial coordinates determined separately
at SPM first and second epoch.
Tycho-2 stars are used to determine the polynomial plate-model coefficients.
In theory, this will yield proper motions on the system of the ICRS.
However, the number of high-quality Tycho-2 stars per field is insufficient
to adequately determine the plate-model coefficients, producing systematic
errors in the proper motions, (i.e. modeling errors), as a function
of position.
These are removed by applying field-dependent adjustments calculated
from the thousands of SPM2 stars per field, effectively transforming the
proper motions to the (differentially determined) system of the SPM2.

As pointed out in Table 3 of Paper II, there were small but significant
differences between the absolute proper motions of SPM1 and the Hipparcos
Catalog.
These we believe can be attributed to residual magnitude equation in the
``motions'' of the galaxies used to set the proper-motion zero point of
the SPM1 and SPM2 Catalogs.
The galaxy image profiles, of course, differ from the stellar profiles upon
which the magnitude equation corrections are based.
An attempt was made to account for the expected difference in the
form of the galaxies' magnitude equation correction, (see Paper I), but some
residual error no doubt remained.
For this reason, we have chosen to use Hipparcos Catalog stars instead
to set the zero point of the absolute proper motions in the SPM3.
The following is a more detailed outline of the SPM3 Catalog's astrometry
reduction pipeline.

The first step in the astrometry reduction is a pre-correction for
differential refraction, across the 6.3 $\degr$ field of view.
Differential color refraction correction is {\it not} deemed necessary,
as the plates are exposed centered on the meridian.

The next step is to solve for the magnitude equation correction, and
simultaneously transform the grating images onto the system of the
central-order images.
The procedure used is identical to that described in Paper I, using central
and first-order image pairs to derive the coefficients of a polynomial
magnitude-equation correction, although in this instance, we have
confined the function to quadratic field terms plus a quadratic magnitude
term.
\be
   \Delta X = F(m, X, Y), ~~~~~\Delta Y = G(m, X, Y)
\label{eq5}
\ee
where
\be
\begin{array}{lrlr}
   F = f_1 m + f_2 mX + f_3 mY \\
\indent + f_4 mX^2 + f_5 mXY + f_6 mY^2 + f_7 m^2, \\
   G = g_1 m + g_2 mX + g_3 mY \\
\indent + g_4 mX^2 + g_5 mXY + g_6 mY^2 + g_7 m^2
\end{array}
\label{eq6}
\ee

In the above, linearized instrumental magnitudes are used, i.e., after
application of the photometric coefficients derived in Equations 2 and 3.
The coefficients, $f_k$ and $g_k$, are found by least squares minimization
of the differences in position between the central order and the average
of the first-order pair of images, for all stars for which these three
images can be measured.
\be
\begin{array}{lrlr}
  X_{(n=1)} - X_{(n=0)} = \\ 
\indent F(m_{(n=1)}, X, Y) - F(m_{(n=0)}, X, Y)
\\
  Y_{(n=1)} - Y_{(n=0)} = \\
\indent G(m_{(n=1)}, X, Y) - G(m_{(n=0)}, X, Y)
\end{array}
\label{eq7}
\ee

Separate sets of coefficients are derived for the
long and short-exposure systems, with typically thousands of stars used
in the long-exposure solution and hundreds in the short.
These are then applied to all images in each exposure system, central through
third diffraction order.

We note that the same magnitude-equation correction has been applied
regardless of whether the images are of stars or galaxies, while the
correction is derived entirely from stellar images.
For this reason, the proper motions of those objects identified as
extended sources in the SPM3 should be used with caution.

After application of the magnitude-equation correction and averaging the
positions of grating pairs, all images are on one of two systems; the 
long-exposure system, or the short-exposure system.
These two are combined using yet another polynomial transformation, this
time a general cubic model.
The number of stars used to determine this ``bridge'' solution was on
average 6500 per field.

With all $X, Y$ measures now on a single system, the transformation
to celestial coordinates can be calculated in a straightforward manner.
Tycho-2 stars, trimmed somewhat by positional and proper-motion error, are
used as reference stars.
These are updated to the epoch of each SPM plate and then transformed to 
standard coordinates using the nominal field centers.
As demonstrated by Platais et al.~(1995), cubic field terms are necessary
to properly describe the SPM plates.
Thus, a general third-order polynomial transformation between plate
coordinates and standard coordinates was assumed, i.e., a 10-term solution
in each coordinate.
The number of Tycho-2 reference stars varied from 170 to 1100, with an average
of 400 per field.
The standard error of unit weight per coordinate ranged from 65 to 200 mas,
with a mode of 80 mas.

For stars with multiple grating and/or exposure-system measures, a single
position is determined from the weighted average of the individual 
derived positions.
There is simultaneously a trimming of discordant position estimates.
The weighting and trimming procedures, as well as the procedure for
estimating internal uncertainties, are similar to those used in
previous SPM reductions, as described in Paper II.
This yields positions, in general, from both the blue and visual plates
at first and second epochs.
Within each epoch, the blue- and visual-derived positions are averaged,
weighted by the number of images per plate for each star.

In theory, having positions on the system of the ICRS at two epochs, one
need only take differences and divide by the epoch difference to derive
absolute proper motions.
This we do, but the resulting proper motions are not considered final.
In practice, the high-order polynomial transformation and the often
insufficient number of Tycho-2 reference stars, can result in significant
systematic effects in the derived positions and proper motions.

The top two panels of
Fig.~2 show differences between the so-derived preliminary positions
and proper motions compared to those of the SPM2
Catalog, as a function of plate coordinate, for a sample SPM field.
We show only the right ascension components of the position and proper-motion
differences as a function of $x$ coordinate, but the trends in the declination
component and as a function of $y$ coordinate are similar in appearance.
The position-dependent effects seen are not unexpected.
The proper motions of the SPM2 were derived in a differential manner,
using a large number of anonymous faint stars to determine the {\it difference}
between the first and second-epoch plates' models.
Thus, they are better determined than these preliminary SPM3 proper motions.
As for the trends in positional differences, these are likely due to
statistical uncertainty in the polynomial coefficients derived from different
sets of reference stars; Tycho stars in the case of SPM2, and Tycho-2 stars
in the case of SPM3.
In both cases, the number of reference stars is only marginally sufficient
to determine the polynomial plate models.
Fortunately, the large number of SPM2 stars in the SPM3 sample allows us
to calibrate the preliminary SPM3 data onto the system of the SPM2.

The very differences shown in Fig.~2 are used to perform this
calibration.
For both the SPM3 positions and proper motions, separately, a correction
is determined based on the median difference with the corresponding SPM2
data.
The 50 SPM2 stars nearest 
to each target SPM3 star 
(spatially, by 2-dimensional separation) 
are used in determining the median correction vectors.
The medians are calculated and applied star by star,
one SPM field at a time, i.e., the corrections vary from field to field.
The top two panels of
Fig.~3 show the differences with SPM2 {\it after} the corrections have
been applied to the SPM3 astrometry.

At this stage, the SPM3 positions and proper motions are on the system
of the SPM2.
The positions in the SPM2 itself, however, are known to suffer from
a systematic field error when compared to Tycho-2 positions.
This was attributed to the inadequacy of the third-order polynomial
used to transform the SPM plate measures into celestial coordinates,
during construction of the SPM2 Catalog.
The SPM3 positions, being on the system of the SPM2, are expected
to suffer similarly and, thus, we shall make an explicit correction
for this systematic field pattern.
The SPM2 proper motions, being differential, should not suffer from
any such effects, provided the telescope and plates remained unchanged
between the two epochs of observation.
Prompted by the referee of an earlier draft of this paper that described
an earlier version of the SPM3 Catalog, we have investigated the
proper motions of the SPM3 for any such field pattern as well.
We were surprised to find that one exists.
It implies that a systematic change has occurred in either the telescope
or the stored plates, causing a difference in the form of the
transformation from sky coordinates into plate coordinates between
first and second epoch.
Whatever its cause, the systematic effect in proper motions must be
corrected for as well.

The corrections, made separately for the positions and proper motions, are
based on residual ``masks''.
In the case of the proper motions, residuals are formed between the
Tycho-2 values and the SPM3 proper motions.
The differences are binned as a function of position within each SPM field, 
(i.e., by plate position), onto a 21$\times$21 grid.
The mean proper-motion differences, upon stacking the 156 SPM fields,
are shown in Fig.~4a.
Bi-linear interpolation within the grid is used to calculate the 
correction to be applied to each object in the SPM3 Catalog.

In the case of the positional correction mask, residuals are formed using
the UCAC 0.9 Catalog and the SPM3.
The much higher density of UCAC over Tycho-2 provides a better determination
of the position vector residuals.
The mean position differences are shown in Fig.~4b.
In calculating the residuals, the SPM3 proper motions are first 
mask-corrected and then used to bring the SPM3 positions to the UCAC epoch.
As with the proper-motion correction, bi-linear interpolation is used
to calculate individual corrections to be applied.
And as with the proper-motion mask, the mean correction across the entire
field is forced to zero, by means of a constant offset, thus keeping the
field-by-field mean position and mean
proper motion on the same system as the SPM2.
The positions in the SPM2 are on the 1991.25 system of the original Tycho
Catalog.
Thus, the SPM3 positions will be on this same system, which in effect,
is the same as that of Tycho-2.

The proper-motion system of SPM2 is independent of Hipparcos and Tycho-2.
As stated earlier, and detailed in Paper II,
the proper-motion zero point of each SPM2 field was
determined from the mean relative motion of the galaxies in that field.
The magnitude equation affecting the galaxy images is expected to be
different in form from that affecting the stellar images, upon which
the corrections are based.
This was accounted for, to first-order, in earlier SPM reductions by 
solving for an effective magnitude offset between the stellar and galactic
magnitude equation corrections, (see Paper I).

Realizing this simple compensation may be inadequate, especially for the
PMM-measured galaxies, we have decided to calculate the absolute
proper-motion zero point of the SPM3 from Hipparcos Catalog stars, and their 
Hipparcos-measured absolute proper motions.
For each SPM3 field, the correction to absolute proper motion
is the median of the differences in proper motion between
the Hipparcos values and the SPM3 measures, after being calibrated to
the SPM2 system.
We note that the proper-motion system of Hipparcos is less likely to
be affected by uncorrected magnitude equation compared to that of Tycho-2,
which relies on ground-based photographic material.
For this reason, we choose to use Hipparcos to determine the absolute
zero-point corrections of the SPM3 proper motions, field by field.

Fig.~5 shows the distribution of the corrections for the 156 fields
that comprise the SPM3.
Each point represents the correction applied to a particular field.
The mean correction is effectively zero in 
$ \mu_{{\alpha}^*} = \mu_{\alpha}cos \delta$,
but in $\mu_{\delta}$ it is +0.55 $\pm$ 0.12 mas~yr$^{-1}$.
This is similar to what was found in a comparison of SPM1 proper motions
with those of Hipparcos, described in Paper II.
The number of Hipparcos stars per field varies from 63 to 147, with an
average of 102.
As will be shown in Section 4, the precision of the SPM3 proper motions
at these magnitudes is roughly 4 mas~yr$^{-1}$.
Thus, the accuracy to which the final SPM3 proper motions are on the
absolute system of Hipparcos is expected to be 0.4 mas~yr$^{-1}$.

\subsection{Compilation}

At the completion of the procedures described above, calibrated B, V
photometry, celestial coordinates, and absolute proper motions will
have been determined for all objects in the input list, for each of the
156 SPM3 fields.
What remains is to combine the data from all fields.
First, though, a culling is made of improperly identified objects.

The SPM3's PMM-measured detections were identified by simple positional
coincidence with the four input catalogs, dominated by 
the 2MASS point source and extended source catalogs.
With no previously determined proper motions for the vast majority of
the target objects, a relatively large positional coincidence tolerance
of 5.5 arcsec was used to match target objects with PMM detections.
This led to false identifications in some cases, in one or both epochs,
due to the large number of spurious detections in the raw PMM measures.

Therefore, a final check of positional coincidence is made, this time
at epoch 2000.
All SPM3 objects are updated to this epoch, using the SPM3 proper motions,
and compared with their input catalog positions.
A matching tolerance of 2 arcsec is enforced.
Roughly 8\% of the tentatively identified objects were determined to be
spurious and eliminated.

The final step is the merging of the 156 fields, averaging the measures
of objects in the overlap area of adjacent fields.
Again, a 2-arcsec positional matching tolerance is used for this purpose.
Averages are weighted, based on the estimated uncertainties in the quantities
being averaged.

\section{Catalog Properties}


The SPM3 Catalog includes 10,764,865 objects.
It contains positions and absolute proper motions, on the system of the ICRS,
and photographic $B, V$ photometry, and estimated uncertainties for each
of these quantities.
The positions are given at epoch 1991.25, the mean epoch of the Hipparcos
and Tycho Catalogs and very nearly the mean epoch of our second-epoch
plate material.
The positions and proper motions have been used to make a cross-referencing
to other selected catalogs, including the original input source catalogs,
by positional coincidence.
This provides both complementary data, e.g. Hipparcos parallaxes,
and comparable data, e.g. Tycho-2 proper motions.
Table 1 summarizes the number of objects in the SPM3 which have been
cross-identified in other catalogs.
The cross-identification data are available along with the SPM3 Catalog
via the web, at http://www.astro.yale.edu/astrom/.

The uncertainties in the SPM3 Catalog are derived from multiple measures
of the same object, in different exposure systems and/or diffraction orders 
and/or passbands and/or fields, for stars in the overlap region of adjacent
SPM fields.
For those objects with only one image and one measurement per plate, 
a value is assigned
based on the mean uncertainty of other (multiply measured) images with similar
signal-to-noise.
Fig.~6 illustrates the Catalog uncertainties as a function of magnitude.
Only uncertainties along right ascension are shown, to save space.
The uncertainties along declination exhibit similar trends.
The uncertainty estimates are quantized, particularly the photometric
estimates.
For the purpose of plotting, the uncertainties shown in Fig.~6 have been
smoothed slightly.
The median uncertainty, as a function of magnitude, is shown for the
positions and proper motions.
The photometric uncertainties are severely quantized due to the
adoption of a baselevel uncertainty of 0.25 mag early in the reduction
procedure, during the magnitude linearization.
Consequently, displaying the median of the photometric uncertainties is
not instructive.
Instead, a comparison is made with the CCD photometry of over 7000 stars
used in the calibration of the photographic photometry.
The dotted curve in the bottom panel of Fig.~6 shows the dispersion
of the differences with this CCD photometry.
Also, an external error estimate in proper motion is made at the 
bright and faint ends of the Catalog distribution.

Based on approximately 11,000 Hipparcos stars, the internal proper-motion
uncertainties underestimate the external values by about 30 percent.
At the faint end, the dispersion in proper motion of roughly 300 quasars,
(originally identified in the SPM2),
is 12 mas yr$^{-1}$, significantly larger than the internal error estimate.
No external comparison is made of the SPM3 positions, but, as with the
SPM2 Catalog, the internal uncertainty estimates most likely also underestimate
the external values by about 30 percent.
Thus, we estimate the positional precision to be 40 mas and the proper-motion
precision to be 4 mas yr$^{-1}$, for well-measured stars in the SPM3.
With approximately 100 Hipparcos stars per field available to set the
proper-motion zero point, we estimate the accuracy of the SPM3 absolute 
proper-motion system to be 0.4 mas yr$^{-1}$.
The photometric precision is $\sim$0.15 magnitudes, for the best-measured
stars.

The input catalog from which the list of SPM3 objects is derived is expected
to be nearly complete to the limiting magnitude of
the 2MASS Point Source Catalog.
Of course, near this limit, the actual completeness of the SPM3 will be
lower.
Issues of image confusion and PMM detection idiosyncrasies affect
the catalog completeness at all magnitudes, but will be most severe
for the faintest stars.
Fig.~7 indicates the magnitude distribution of the final SPM3 Catalog.
For comparison, also shown are the distributions of other catalogs over 
the same area of sky, including the PDS-based SPM2 Catalog, which was
not designed to be complete to any magnitude.
The falloff of the SPM3 distribution beyond $V$=17.5 is
remarkably abrupt, given that it must reflect both the variation in
plate depth among the 156 fields and a broadening by the
magnitude uncertainties.

It is not a trivial task to quantify the completeness of the SPM3
as a function of magnitude without an external catalog that is known to 
be 100\% complete to a deeper limit and in a comparable passband.
However, we can make an approximate estimate of the 
relative completeness between the SPM3 and the 2MASS point source catalog.
In order to make a direct comparison, as a function of SPM3 magnitude,
we require an approximate transformation from 2MASS $J, H, K$ magnitudes
to $V$.
We use a relation, that we have determined empirically,
\be
  V \approx J + 2.79 (J-K)
\label{eq8}
\ee
and find to be an adequate approximation for a large range of stellar types.
Applying this relation, we calculate and display in Fig.~8 the 
$V$-magnitude distribution of the 2MASS point source catalog and compare it to
that of the SPM3, in a sample SPM field.
In the range $8<V<17.5$, the SPM3 distribution is similar to that of 2MASS,
to within the statistical noise.
The relative completeness drops from 1.03 at $V$=17.5 to 0.56 at $V$=18.0.
Thus, the completeness magnitude limit of the SPM3 Catalog is taken to be
$V$=17.5.

Even for stars brighter than this magnitude, image confusion and 
blending cause a low level of incompleteness.
At the very bright end of the magnitude distribution, this incompleteness can
be estimated using the Hipparcos Catalog.
All Hipparcos stars were included in the input target list and yet the
final number of Hipparcos stars in the SPM3 Catalog, 10,900, is 8\% below
the expected number of $\sim$ 11,790.
By comparison, the SPM2 Catalog, based on PDS measurements and a far
less automated reduction pipeline, contains 10,957 Hipparcos stars.
A large fraction of these missing Hipparcos stars are brighter than $V$=5,
and not measureable on the SPM plates.

Perhaps the best estimate of the fraction of stars lost because of the 
more automated SPM3 reduction pipeline, and other factors, 
can be made using the SPM2
as a comparison.
Assuming a conservative positional matching tolerance of 2 arcsec, we find
6\% of SPM2 stars and galaxies are absent from SPM3.
Assuming a more liberal 3-arcsec tolerance, the percentage of absentees
falls to 4\%.
Of these missing stars and galaxies, there is little dependence on magnitude.
However, a significant fraction of the absent stars have large proper motions,
based on the SPM2 measures.
For instance, of the 14,193 absentees (based on the 3-arcsec matching),
1939 are stars with total proper motions in excess of 180 mas~yr$^{-1}$.
This benchmark value of 180 mas~yr$^{-1}$ is relevant given the $\sim$30-year
mean difference between the first-epoch SPM plates and the 2MASS Catalog,
(the dominant input catalog), and the matching tolerance of 5.5 arcsec.
Stars with proper motions in excess of this value, and not in the Hipparcos
or Tycho-2 Catalogs, would not have been properly identified and, thus,
are absent from the SPM3.
An effort was made to include these high proper-motion stars in
the construction of the SPM2, but for the sake of expediency, they were
not specifically included in the SPM3.
Thus, we caution the potential user of the SPM3 Catalog as to the severe
incompleteness of faint, high-proper motion stars.

\section{Catalog Versions}

With several versions and subversions of the SPM Catalog in existence,
it is prudent to describe these briefly, for the sake of clarity.
The original SPM 1.0 is based on PDS measures of 30 SPM fields around the
South Galactic Pole.
It covers roughly 720 square degrees and contains 58,880 objects.
The input catalog for the SPM 1.0 is comprised of objects of interest gleaned
from the literature, utility objects (i.e., stars necessary to the
reduction procedure), visually confirmed galaxies for proper-motion
reference, and a large number of randomly-selected anonymous stars intended
for kinematic study.
The SPM 1.0 is not designed to be complete over any magnitude range.

The SPM 2.0 is based on a re-reduction of the measures used to construct
the SPM 1.0, in addition to PDS measures of other mid-declination SPM fields
for which first and second-epoch plate material exists.
It includes measures from a total of 156 SPM fields, covering an area of
approximately 3700 square degrees.
The fields are from the -25, -30, -35, and -40 degree declination bands,
excluding fields near the Galactic Plane.
The input catalog for this version of the SPM Catalog was constructed in
a similar fashion to that of the SPM 1.0.
Thus, the SPM 2.0 is also not explicitly designed to be complete over
any magnitude range.
The SPM 2.0 contains 321,608 objects.

The reduction procedure for the SPM 2.0 incorporates one significant
improvement over that of the SPM 1.0, that being the handling of the
magnitude-equation correction for galaxies.
As an aid to those who had made kinematic studies using the SPM 1.0, the
SPM 1.1 was created by extracting all objects from the SPM 2.0 that fell
within the boundary of sky defined by the SPM 1.0.
A net gain of 8 objects resulted, yielding 58,887 entries in the SPM 1.1.
The SPM 1.1 is a subset of the SPM 2.0.
The identification numbers of objects in the SPM 1.0 and SPM 1.1 are not
the same.

The position system of the SPM 1.0, 1.1 and 2.0 is that of the original Tycho
Catalog, at epoch 1991.25.
The absolute proper-motion system of these versions is
based on PDS measures of galaxies, on a field by field basis.
The SPM 1.0 is described in Paper II, (Platais et al.~1998).
Details concerning the SPM 1.1 and SPM 2.0 can be found at
http://www.astro.yale.edu/astrom/

The SPM 3 is constructed from PMM measures of the same 156 SPM fields which
comprised the SPM 2.0 Catalog.
Actually, PMM measures were made of all existing first and second-epoch SPM
plates.
As part of a project to construct an input catalog for the FAME mission, a
reduction pipeline for these PMM data was created at the US Naval Observatory.
This pipeline transformed the PMM measures into celestial coordinates
at plate epoch, and incorporated many of the features of the reduction
procedure of the earlier SPM Catalogs.
The input catalog for this pipeline was much more complete, though.
It consisted of the union of the Hipparcos, Tycho-2, and UCAC 0.9 Catalogs.
Later, in order to reach the SPM plate limit, a magnitude-trimmed version
of the USNO-A2 Catalog was added to the list of input source catalogs.

Having this reduction pipeline in place, it was realized that it could be
used to create a new version of the SPM Catalog, one nearly complete to
the SPM plate limit.
Thus, the reduction pipeline was modified slightly to allow it to be
applied to the PMM measures of both first and second-epoch SPM plates.
(The original US Naval Observatory project made use of only the first-epoch
SPM material.)
With first and second-epoch celestial coordinates, preliminary proper
motions could then be derived, and then transformed to a better-calibrated
system, as described in detail elsewhere in this paper.

The SPM 3.0 was the first version to be based on PMM measures in this way.
It was made avaliable for only a short time, and directly distributed to
only a handful of colleagues.
Upon the release of the full-sky 2MASS data, it was decided to replace
the USNO-A2 as an input source catalog with the 2MASS Point Source Catalog
(which does include the 2MASS extended sources).
This resulted in the SPM 3.1 Catalog, with improved completeness properties,
along with the added bonus of cross-identification to the 2MASS photometry.
This version of the SPM Catalog was also available only for a brief
period and once again directly distributed to only a handful of researchers.

It was the comments of the referee on an earlier version of this paper
which led us to discover the systematic pattern in the stacked proper-motion
residuals of the SPM 3.1.
As described in Section 3, this systematic error now has been corrected,
resulting in the current version of the catalog presented here, SPM 3.2.
The SPM 3.2 supercedes both the SPM 3.0 and SPM 3.1.
It differs from the previous major version of the SPM Catalog,
the SPM 2.0, in two significant ways:
The use of PMM measures, as opposed to PDS measures, results in a slightly
degraded precision, but a far greater completeness in the SPM 3.2.
And the use of Hipparcos star proper motions, as opposed to external
galaxies, places the SPM 3.2 on the ICRS proper-motion system.

\section{Kinematics of Metal-poor, Main-sequence A Stars}


In a spectrophotometric survey at the South
Galactic pole (SGP), Rodgers, Harding and Sadler (1981, hereafter RHS81)
identified a group of
A-type stars located between 1 and 4 kpc from the Galactic
plane and with main-sequence surface gravities, rather than horizontal-branch
ones, as one might expect in an old-halo or thick-disk population. 
Lance (1988a,b, hereafter L88)
increased the sample, and re-determined radial velocities, calcium abundances,
surface gravities, and distances. 
She confirmed the previous results of RHS81. 
The calcium abundance of the main-sequence stars in Lance's sample
ranges between 0 and -0.6 dex, and the radial velocity dispersion is
62 km~s$^{-1}$, somewhat higher than the vertical velocity dispersion of the
thick disk ($\sim$ 40 km~s$^{-1}$, e.g., Edvardsson et al.~1993).
On these grounds,
L88 and RHS81 hypothesized that these stars were formed in a
system accreted by the Milky Way.
Rodgers and Roberts (1993) expanded this type of survey to two additional
Galactic locations, in order to determine the rotational
properties of this population from radial velocities alone.
They found that the kinematical properties of the A-type,
main-sequence stars are similar to those of the thick disk. 
Thus, a new hypothesis was put forward:
the stars in question may be thick-disk blue stragglers. 
However, Rodgers and Roberts (1993) concluded this was unlikely, 
estimating that about a third of the thick disk stars would have to 
be blue stragglers,
a number that is too large when compared with the frequency of
blue stragglers in globular clusters.

We have combined the radial-velocity data from L88 with SPM3
proper motions to determine 3D velocities of 29 A-type, main sequence
stars at the SGP.
There was no overlap between the Rodgers and Roberts (1993)
sample and the SPM3 Catalog.
For comparison, we
also examine the A-type, main-sequence stars identified in
Wilhelm et al.~(1999, hereafter W99), Table 2C, a
catalog of metal-poor stars.
There are 79 stars in common between the SPM3 Catalog and the W99 data set.
This latter sample
has a metallicity range from -0.1 to -3.0 dex, much wider
than that of L88.
Thus, the W99 sample is expected to contain a significant fraction of
halo stars.

We have determined velocity components in a
cylindrical coordinate system, in the Galactic rest frame.
$\Pi$ is the radial component
(positive outward from the Galactic center), $\Theta$ is the rotational
component (positive toward Galactic rotation), and $W$
is perpendicular to the Galactic disk
(positive toward the North Galactic Pole).
In Fig.~9 we show the velocity components
of the L88 sample, while in Fig.~10 those of W99.
The majority of the L88 stars have velocities typical of the
thick disk, ($\Theta \approx 180$ km~s$^{-1}$),
while the W99 velocities are characteristic of both the thick
disk and halo.
Roughly half of the stars in this latter sample
have metallicities lower than -1.0 dex, the traditional cutoff between
thick disk and halo.

The proper-motion results for the L88 sample, thus, support the case for
thick-disk blue stragglers, rather than that for an accreted component
with distinct kinematics.
More recently, in a radial-velocity
study of blue metal-poor stars, Preston and Sneden (2000) (also supported
by Carney et al.~2001)
found a high fraction of binaries, leading them to suggest that
the frequency of blue stragglers in the halo (i.e., field stars) is
larger than that in globular clusters.
Their findings, as do ours, lend support to the notion that
distant, A-type, main-sequence stars are more likely Galactic
field blue stragglers
rather than being the remnants of a disrupted satellite.

\section{Summary}


The recently released SPM3 Catalog provides positions, absolute proper motions,
and photographic $B, V$ photometry for 10.7 million objects, primarily stars,
in an irregular declination band in the Southern hemisphere.
Unlike previous versions of the SPM Catalog, we strive for completeness
to the limiting magnitude of the plate material.
This is attained by using the Precision Measuring Machine of the
the US Naval Observatory to scan the first and second-epoch plates.
The resulting catalog is nearly complete in the range $5<V<17.5$.
Also, this is the first SPM Catalog in which the zero-point of the
absolute proper-motion system is defined by Hipparcos stars, as opposed
to external galaxies.
As with all previous versions of the Catalog, a correction for
magnitude equation is derived and applied to each individual plate.
The statistics of the SPM3 Catalog are summarized in Table 1.
The Catalog, along with cross-identifications to other relevant catalogs,
is available via electronic transfer.

As a brief example of the application of the SPM3 proper motions,
we calculate 3D velocities for metal-poor, main-sequence A stars, a group
first identified by Rodgers (1971) and subsequently expanded upon.
We find these stars' kinematics are predominantly similar to that of
the thick-disk, lending support for their interpretation as thick-disk
blue stragglers, as opposed to being the remnants of an accretion process.
This is but a single, cursory example from among
the large number of kinematic studies possible with the SPM3 data, which
are currently being pursued.

Future versions of the SPM Catalog, expanding coverage to a larger portion
of the Southern sky, will be based on the existing first-epoch photographic
plate material and second-epoch CCD observations now underway.
A CCD camera system is currently operating on the double astrograph of 
the Cesco Observatory and providing the necessary second-epoch SPM material.
The system consists of an astrometric/photometric 4K$\times$4K PixelVision
camera mounted in the visual focal plane and a photometric 1K$\times$1K Apogee
camera in the blue focal plane.
Specific regions of interest, e.g. globular-cluster fields and Milky Way
satellite fields, are expected to be the first CCD-based SPM projects,
with the continuation of the survey portion of the Southern Proper Motion
program to follow.


The SPM3 Catalog would not be possible without the efforts of the
numerous individuals who made the first and second-epoch observations
at Cesco Observatory,
and those who performed the the PMM scanning at USNO-Flagstaff.
We are grateful to them all.
We also are grateful for the collaborative efforts of Tim Beers, who
prompted us to construct this version of the Catalog, while awaiting the
accumulation of CCD data necessary for future versions.
This project has also benefited from the assistance and helpful comments
of Norbert Zacharias.
Finally, insightful comments from Bob Hanson, who acted as referee to an
earlier version of this paper, were extremely helpful in improving not only
this paper, but the SPM3 Catalog itself.
This publication makes use of data from the Two Micron All Sky Survey, (2MASS),
which is a joint project of the University of Massachusetts and the Infrared
Processing and Analysis Center/California Institute of Technology, funded 
by the National Aeronautics and Space Administration and the National Science
Foundation.
The Southern Proper Motion program has been funded over the years by a
number of grants from the National Science Foundation, including the most
recent, AST-0098548.

\clearpage

\begin{table*}
\begin{center}
\begin{tabular}{lr}
\multicolumn{2}{l}{Table 1.  SPM3 Statistics.} \\
\tableline
\tableline
Sky coverage (sq. deg.)     &                   $\sim 3700$  \\
Magnitude range             &  $4 \lesssim V \lesssim 17.5$  \\
Total no.~of SPM3 objects   &                    10,764,865  \\
Number common to...         &                                \\
~SPM2                     &                       301,272  \\
~2MASS point sources      &                    10,619,404  \\
~2MASS extended sources   &                       113,510  \\
~Tycho-2                  &                       166,757  \\
~Hipparcos                &                        10,901  \\
\tableline
\end{tabular}
\end{center}
\end{table*}


\begin{figure*}
\epsscale{1.6}
\plotone{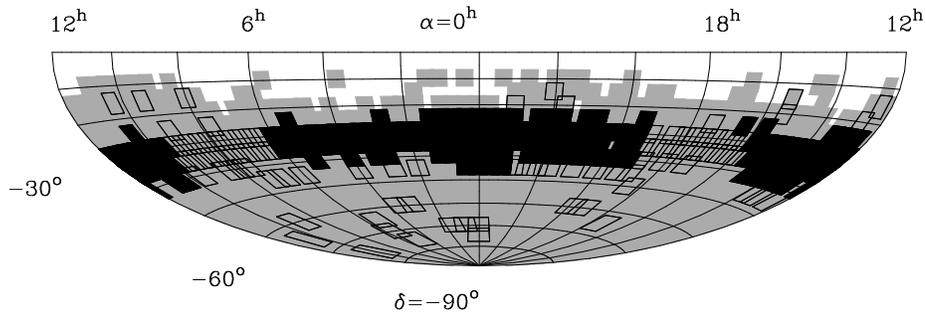}
\caption{
Sky coverage of the SPM, in equatorial coordinates.
The gray area indicates the first-epoch plate coverage.
The dark area shows the extent of the SPM3 Catalog.
Those fields with second-epoch plates, but not included in the SPM3 Catalog,
are indicated by hollow frames.
}
\end{figure*}

\begin{figure*}
\epsscale{1.0}
\plotone{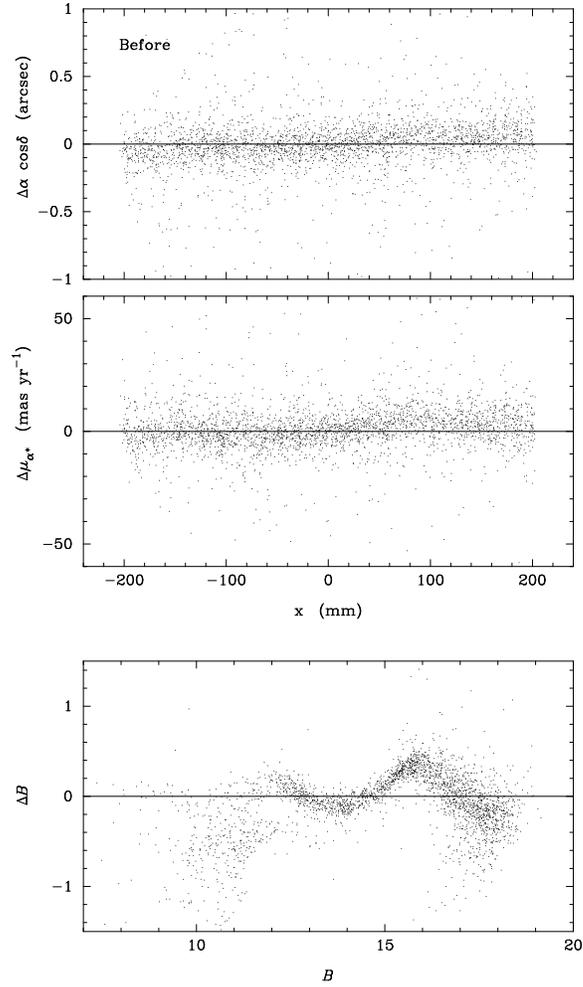}
\caption{
Differences
in positions, proper motions, and photometry,
SPM2 minus preliminary SPM3, before recalibration with the SPM2 data.
Differences along right ascension and in $B$ magnitudes are shown.
The trends in the declination differences and in $V$ magnitudes are of
similar size and appearance.
The sample field shown is \# 333.
}
\end{figure*}

\begin{figure*}
\plotone{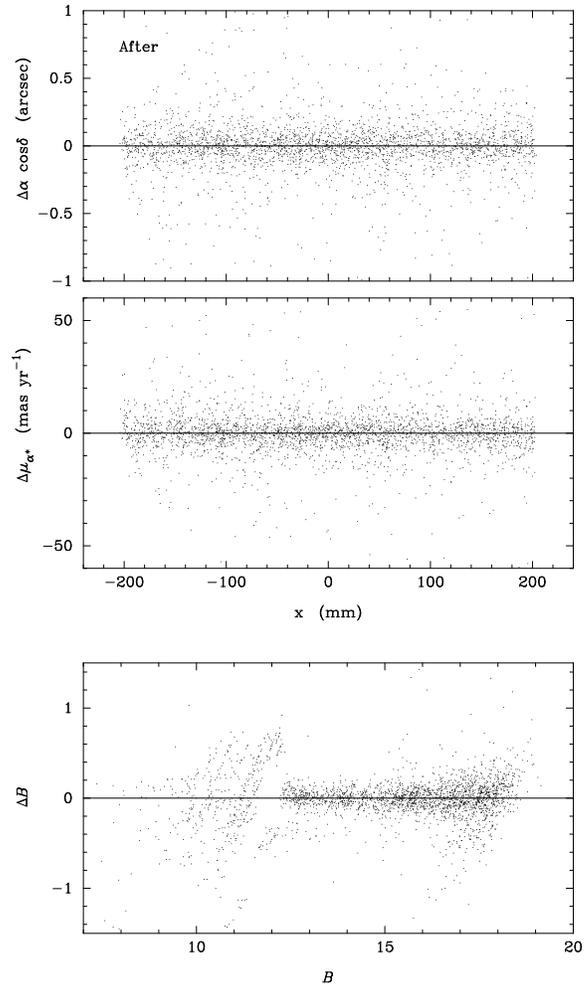}
\caption{
Differences
in positions, proper motions, and photometry,
SPM2 minus SPM3, after recalibration with the SPM2 data.
The sample field is the same as in Fig.~2, \# 333.
}
\end{figure*}

\begin{figure*}
\plotone{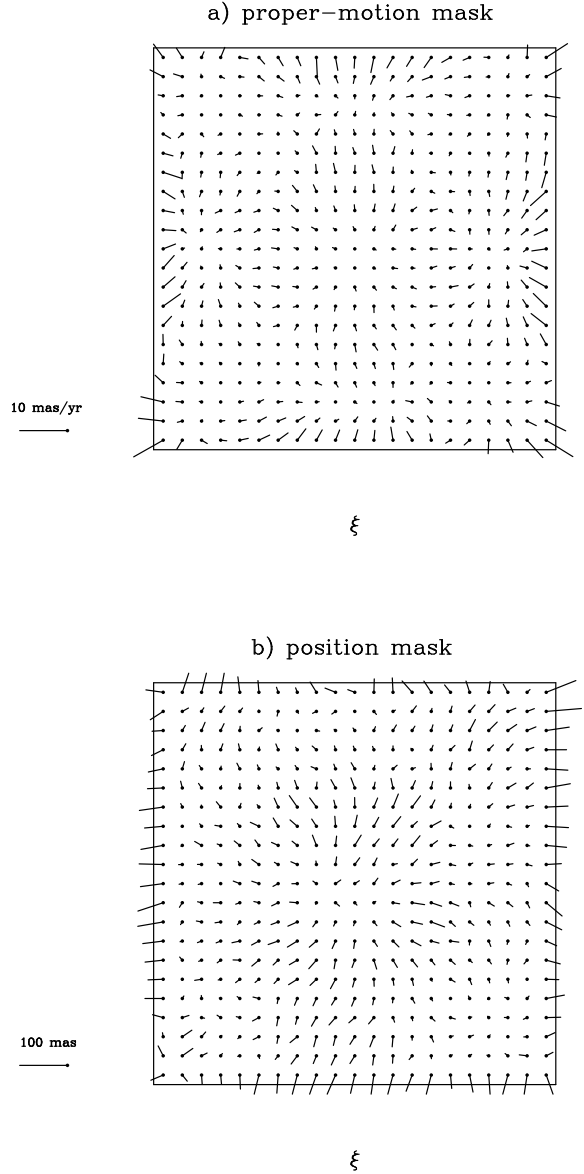}
\caption{
Proper-motion and position correction masks.
a)  Proper-motion residuals, Tycho-2 minus SPM3 (after recalibration
with the SPM2 data), are stacked and averaged as a function of plate
coordinates for all 156 fields to construct the proper-motion correction
mask.
b)  Position residuals, UCAC 0.9 minus SPM3 (after recalibration with the
SPM2 data), are similarly stacked and averaged to construct the
position-correction mask.
The plate boundary indicated is 6.4 degrees on a side.
Proper-motion and position correction vectors of 10 mas~yr$^{-1}$
and 100 mas, respectively, are shown for scale.
}
\end{figure*}

\begin{figure*}
\plotone{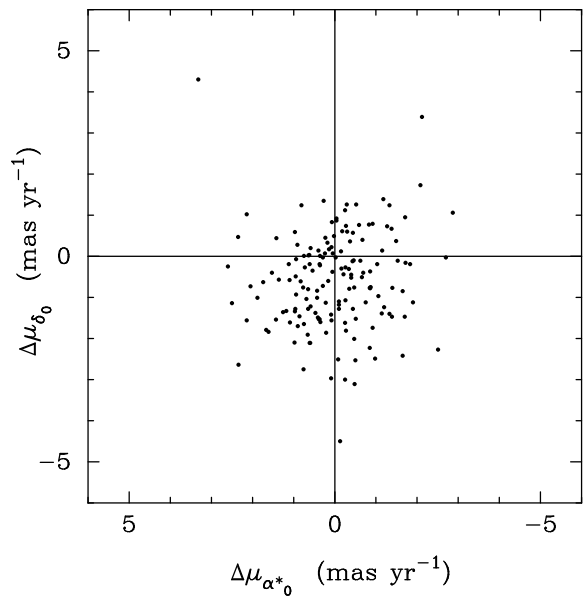}
\caption{
Corrections to absolute proper motion.
The mean difference between Hipparcos proper motion and preliminary
SPM3 proper motions on the system of the SPM2, defines the individual
correction for each field.
The mean correction in ($\mu_{\alpha} cos \delta$, $\mu_{\delta}$)
for the 156 fields is (0.00, -0.55) mas~yr$^{-1}$ with
dispersion (1.12, 1.18) mas~yr$^{-1}$.
}
\end{figure*}

\begin{figure*}
\plotone{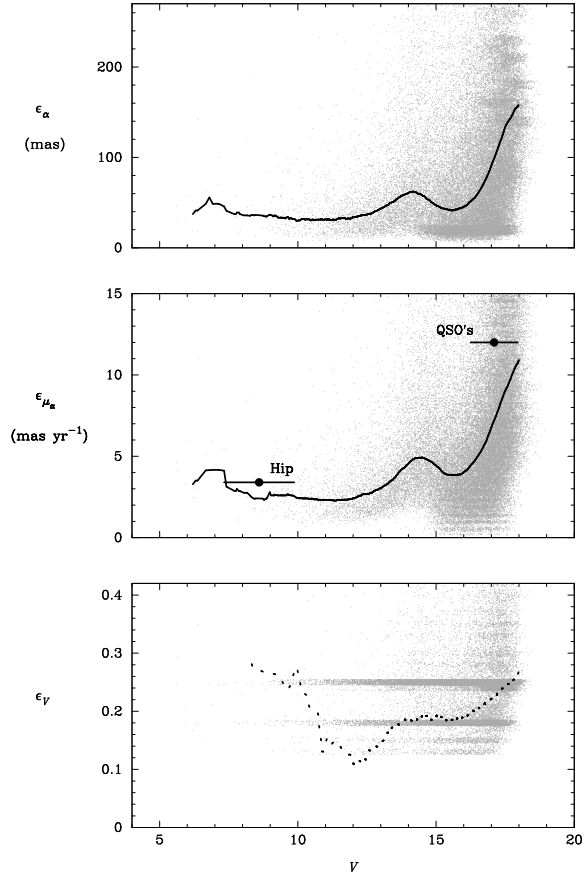}
\caption{
Estimated uncertainties in position (top panel),
proper motion (middle panel), and $V$ magnitude (lower panel).
The points show only a subsample of the Catalog estimated uncertainties,
for the sake of clarity.
In the top two panels, the solid curve shows the median for the entire
Catalog.
Comparisons with the Hipparcos Catalog motions and the proper motions of
identified QSO's give external error estimates, as shown in the middle
panel.
The horizontal bars are indicative of the $V$ dispersion of these two
samples.
The photometry uncertainties are heavily quantized, due to the adoption of
a 0.25-magnitude base-level uncertainty in the photometry linearization
step, early in the photometry reduction procedure.
Thus, a median curve is not shown in the lower panel.
Instead, the dotted curve shows the dispersion of differences in $V$
with the CCD-determined calibration photometry.
}
\end{figure*}

\begin{figure*}
\plotone{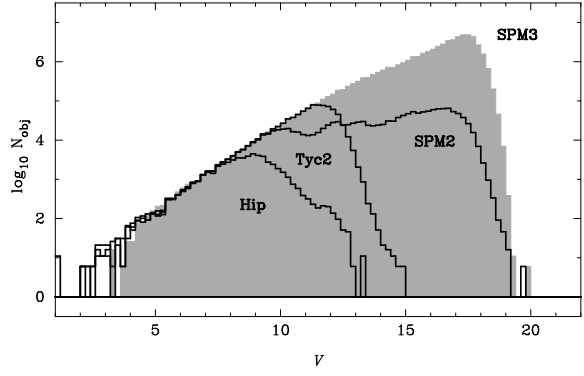}
\caption{
Magnitude distribution of SPM3.
For comparison, the distributions of the Hipparcos Catalog, Tycho-2 Catalog,
and SPM2 Catalog are shown, for the same sky coverage.
}
\end{figure*}

\begin{figure*}
\plotone{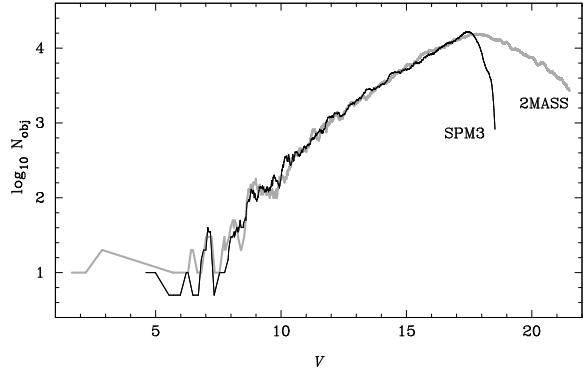}
\caption{
Comparative magnitude distributions for a sample
SPM3 field.
Using an approximate relation for $V$ as a function of 2MASS $J$ and $K$,
the distributions of SPM3 and 2MASS Catalogs can be compared directly.
The completeness of the SPM3 is approximately the same as that of 2MASS
for $V<17.5$.
}
\end{figure*}

\begin{figure*}
\plotone{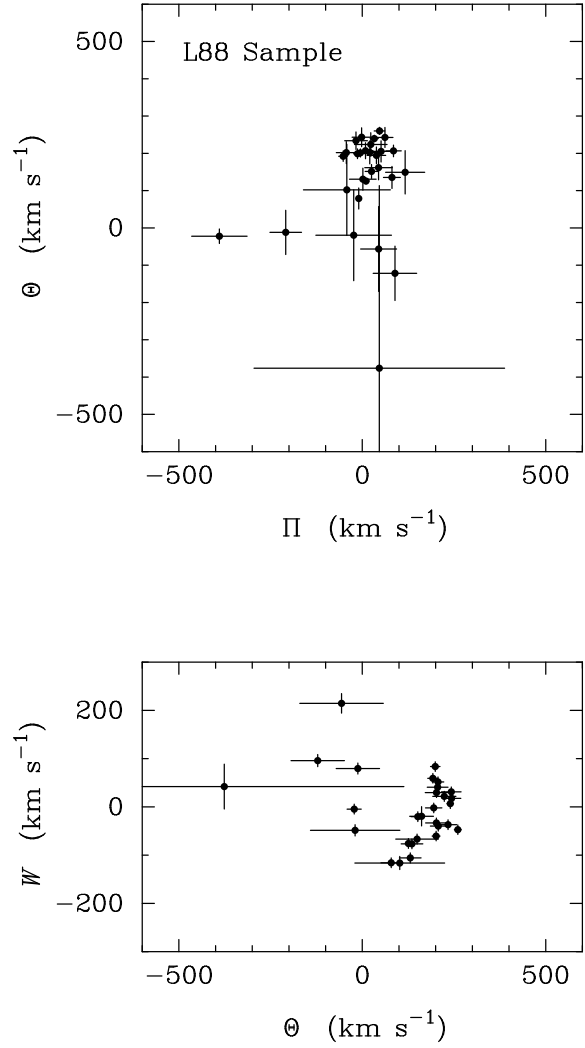}
\caption{
Galactocentric velocities of metal-poor,
main-sequence A stars.
The sample is that of Lance 1988a, 1988b, (L88).
}
\end{figure*}

\begin{figure*}
\plotone{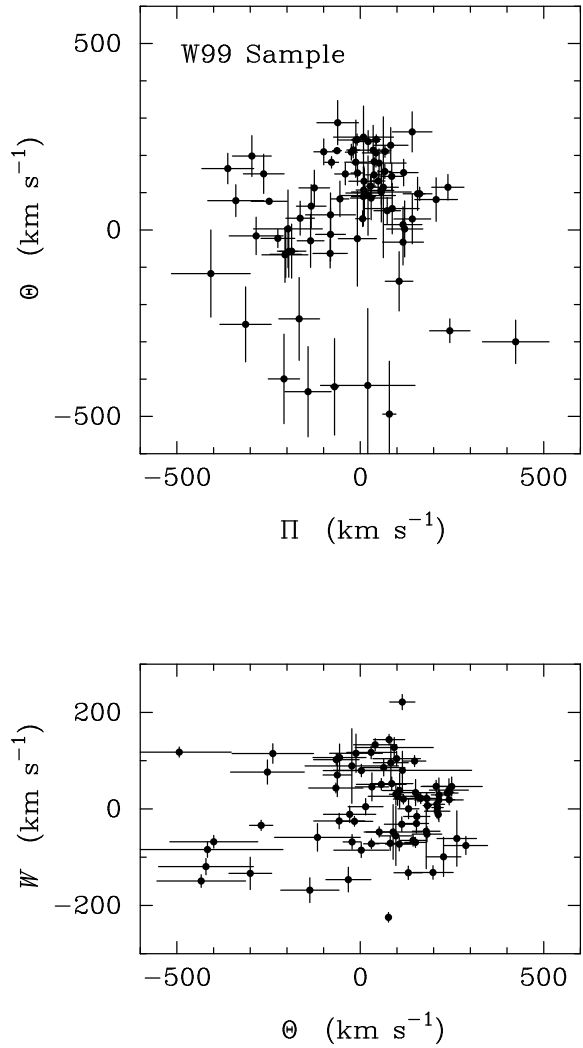}
\caption{
Galactocentric velocities of metal-poor,
main-sequence A stars.
The sample is that of Wilhelm et al. 1999, (W99), and covers a wider range
in metallicity, i.e. extending to lower values, than the L88 sample shown
in Fig.~9.
}
\end{figure*}

\end{document}